\documentclass[aps,pra,reprint,superscriptaddress]{revtex4-2}

\usepackage{amsmath,amsfonts,amssymb}
\usepackage{epsfig}
\usepackage{siunitx}
\usepackage{enumitem}
\usepackage{dcolumn}
\usepackage{bm}
\usepackage{multirow}
\usepackage{xcolor}
\usepackage[hyperindex,breaklinks]{hyperref}
\newcommand{\eref}[1]{Eq.~(\ref{#1})}
\DeclareSIUnit\gauss{G}
\DeclareSIUnit\torr{Torr}

\begin{document}

\title{High-precision measurement and \emph{ab initio} calculation of the $(6s^26p^2)\,^3\!P_0 \rightarrow \, ^3\!P_2$ electric quadrupole transition amplitude in $^{208}$Pb}

\author{Daniel L. Maser}
\affiliation{Department of Physics, Williams College, Williamstown, Massachusetts 01267, USA}
\affiliation{Department of Physics, Astronomy, and Geophysics, Connecticut College, New London, Connecticut 06320, USA}

\author{Eli Hoenig}
\altaffiliation{Current address: Department of Physics, University of Chicago, Chicago, Illinois 60637, USA}
\affiliation{Department of Physics, Williams College, Williamstown, Massachusetts 01267, USA}

\author{B.-Y. Wang}
\altaffiliation{Current address: Department of Physics, Stanford University, Stanford, California 94305, USA}
\affiliation{Department of Physics, Williams College, Williamstown, Massachusetts 01267, USA}

\author{P. M. Rupasinghe}
\altaffiliation{Current address: Department of Physics, SUNY Oswego, Oswego, New York 13126, USA}
\affiliation{Department of Physics, Williams College, Williamstown, Massachusetts 01267, USA}

\author{S. G. Porsev}
\affiliation{Department of Physics and Astronomy, University of Delaware, Newark, Delaware 19716, USA}
\affiliation{Petersburg Nuclear Physics Institute of NRC ``Kurchatov Institute'', Gatchina, Leningrad District 188300, Russia}

\author{M. S. Safronova}
\affiliation{Department of Physics and Astronomy, University of Delaware, Newark, Delaware 19716, USA}
\affiliation{Joint Quantum Institute, National Institute of Standards and Technology and the University of Maryland, Gaithersburg, Maryland 20742, USA}

\author{P. K. Majumder}
\email{pmajumde@williams.edu}
\affiliation{Department of Physics, Williams College, Williamstown, Massachusetts 01267, USA}

\date{\today}

\begin{abstract}
We have completed a measurement of the $(6s^26p^2)\, ^3\!P_0 \rightarrow \, ^3\!P_2$ 939 nm electric quadrupole ($E2$) transition amplitude in atomic lead. Using a Faraday rotation spectroscopy technique and a sensitive polarimeter, we have measured this very weak $E2$ transition for the first time, and determined its amplitude to be  $\langle ^3\!P_2 || Q || ^3\!P_0  \rangle$ = 8.91(9) a.u.. We also present an \emph{ab initio} theoretical calculation of this matrix element, determining its value to be 8.86(5) a.u., which is in excellent agreement with the experimental result. We heat a quartz vapor cell containing $^{208}$Pb to between 800 and \SI{940}{\celsius}, apply a $\sim \! 10 \, {\rm G}$ longitudinal magnetic field, and use polarization modulation/lock-in detection to measure optical rotation amplitudes of order 1~mrad with noise near \SI{1}{\micro\radian}. We compare the Faraday rotation amplitude of the $E2$ transition to that of the $^3\!P_0 -\, ^3\!P_1$ 1279~nm magnetic dipole ($M1$) transition under identical sample conditions.
\end{abstract}


\maketitle

\section{Introduction}
\label{intro}

Atoms have long served as testbeds for precision measurements and low-energy tests of fundamental physics. Searches for new physics, including potential candidate particles for dark matter, are ongoing using, for example, the technology of atomic magnetometers \cite{Budker2014}, atomic clocks \cite{Nicholson2015, Derevianko2014}, and atom interferometers \cite{Hamilton2015}. A comprehensive recent review of the role of atoms and molecules in these searches can be found in \cite{Safronova2018}. 

A particular class of these atomic physics experiments has exploited the symmetry-violating properties of the weak interaction to study atomic parity nonconservation (PNC), and thus potentially probe both electroweak Standard Model physics and potential new physics. A number of these measurements have reached the 1\% level of experimental accuracy \cite{Antypas2019, Wood1997, Vetter1994, Meekhof1993}. Since electroweak effects in neutral atoms scale rapidly with the atomic number, $Z$, such atomic-physics-based tests have focused on heavy atoms, and require independent theoretical wavefunction calculations in the relevant atomic systems to link measured experimental observables to fundamental parameters \cite{Khriplovich}.

Cesium, the heaviest stable alkali element, is an example of an atomic system where very high-precision \textit{ab initio} atomic theory \cite{Porsev2009, Dzuba2012} has come together with precise experimental efforts \cite{Wood1997} to provide an important low-energy test of electroweak physics. More recently, significant progress has been made in \textit{ab initio} calculational techniques for multi-valence atomic systems \cite{Safronova2008, Safronova2009}. In the trivalent thallium system, an existing high-precision PNC measurement \cite{Vetter1994}, coupled with high-precision calculations \cite{Kozlov2001}, has yielded another atomic-physics-based electroweak test. Current theory accuracy lags that of experiment by roughly a factor of two, so that modest further improvements in multi-valence theory will have a significant impact. In a close experiment/theory collaboration, we have completed a series of precise measurements of atomic properties of thallium and its trivalent cousin indium \cite{Ranjit2013, Augenbraun2016, Vilas2018}, which have served as benchmarks for ongoing calculational efforts \cite{Safronova2013}. In particular, by comparing a series of excited-state polarizability measurements in indium to theoretical predictions from two complementary calculational approaches, we were able to show that a configuration interaction (CI) approach, combined with the coupled-cluster (CC), all-orders method to the three-valence system gave better agreement with experiment than the pure CC method \cite{Vilas2018}.

Recently, Porsev \textit{et al.} have undertaken a new \textit{ab initio} calculation of the atomic structure of tetravalent lead \cite{Porsev2016}. Two high-precision parity nonconservation optical rotation experiments were completed in the 1990s \cite{Meekhof1993, Phipp1996}, but the atomic theory accuracy at that time in this complicated system was estimated to have an uncertainty near 10\%, limiting the potential impact of the measurements on testing electroweak parameters. The 2016 theory work \cite{Porsev2016} improves the precision of the PNC calculation by better than a factor of two. Testing the accuracy of this new calculation and guiding forward further improvements will require a similar suite of benchmark measurements in lead. Beyond some energy level measurements and hyperfine structure measurements in $^{207}$Pb \cite{Bouazza2001, Persson2018}, measurements of atomic properties such as transition amplitudes and polarizabilities at the 1\% level of accuracy do not exist for this element.

Here we present a new measurement and accompanying \textit{ab initio} calculation of the lead ground-state $^3\!P_0 \rightarrow \,^3\!P_2$ electric quadrupole ($E2$) transition amplitude. We intend to follow up this result with future measurements of lead excited-state polarizability within the $6s^2 6p 7s$ manifold (see Fig.~\ref{fig_structure}) using similar techniques and apparatus used for our earlier polarizability work in thallium and indium. Thus, we also include relevant \textit{ab initio} polarizability calculations in Sec.~\ref{theory}. 

In the present transition amplitude work, we measure the ratio of the $E2$ amplitude to the that of the ground-state $^3\!P_0 \rightarrow \, ^3\!P_1$ magnetic dipole ($M1$) transition amplitude. This allows us experimentally to eliminate a number of common factors responsible for measured absorptivity of both transitions and extract a ratio of quantum-mechanical amplitudes. Because the $M1$ amplitude is precisely calculable without detailed wavefunction knowledge \cite{Porsev2016}, we ultimately can determine the $E2$ amplitude (proportional to the transition quadrupole moment) from our experimental ratio measurement. Comparative absorptivity measurements have been completed recently \cite{Rafac1998, Antypas2013, Damitz2019} in Cs, producing high-precision determinations of transition amplitude ratios for electric dipole (E1) transitions, but to our knowledge this is the first such measurement using E1-forbidden transitions.

The $E2$ transition linestrength is roughly a factor of 30 weaker than that of the already-weak $M1$ transition. In this work, a highly sensitive optical polarimetry technique \cite{Meekhof1995, Majumder1999, Kerckhoff2005} was used to measure the Faraday rotation signals of the two transitions in an identical longitudinal magnetic field. An analogous precision measurement of the $E2$/$M1$ amplitude ratio within the Tl $6p_{1/2} \rightarrow 6p_{3/2}$ transition was completed in our laboratory using a similar technique some years ago \cite{Majumder1999}. In Sec.~\ref{background}, we outline the atomic structure details involved with extracting transition amplitude information from the observed Faraday rotation lineshapes. Secs.~\ref{experiment} and \ref{results} include a description of the experimental apparatus, method, and data analysis. Sec.~\ref{theory} outlines the \textit{ab initio} theoretical calculation of the electric quadrupole matrix element, and also the atomic polarizability of several relevant excited states of lead. We conclude with a comparison of experiment to theory.

\section{Atomic Structure and Faraday Rotation Lineshape}
\label{background}

\begin{figure}
\includegraphics[width=0.85\columnwidth]{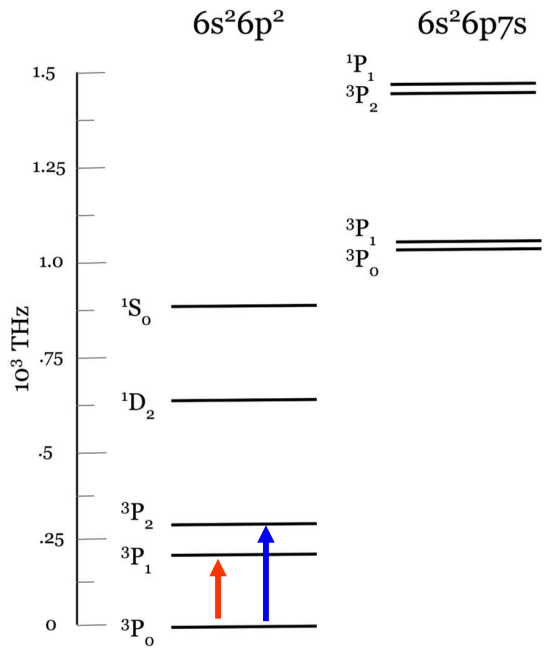}
\caption{Low-lying energy levels of $^{208}$Pb, with the $M1$ and $E2$ transitions shown in red and blue, respectively.}
\label{fig_structure}
\end{figure}

For these spectroscopic studies, we made use of an isotopically enriched (99.9\%) sample of $^{208}$Pb ($I=0$), providing us with a simple, single-feature spectroscopic lineshape for both transitions studied.  Fig. \ref{fig_structure} shows an energy level diagram for the relevant states. Due to the intrinsically weak nature of the $E2$ transition, there is no detectable direct absorption feature, even at the highest sample temperature and density we can achieve. We therefore choose to focus on the real, rather than imaginary, part of the refractive index, and measure the milliradian-sized Faraday rotation lineshape induced by a small longitudinal magnetic field. The observed optical rotation results from the difference in the Zeeman-shifted refractive indices, $n_\pm$, for right and left-circularly-polarized electric field components driving $\Delta m = \pm 1$ transitions originating from the $|^3\!P_0, \, m = 0 \rangle$ ground state. The Faraday rotation signal can be written
\begin{equation}
\label{eq1}
\Phi_F(\omega)=\frac{\omega \ell}{2c} (n_+(\omega) - n_-(\omega)),
\end{equation}
where $\ell$ is the interaction path length through the optically active medium, $\omega$ is the laser frequency, $c$ is the speed of light, and $n_{\pm}$ represents the dispersive real part of the refractive index for a given circular polarization.  

The application of a small magnetic field, ${\bf B} = B_0\hat{z}$, parallel to the laser propagation direction causes equal and opposite Zeeman shifts to the resonant frequency of the circular polarization components, $\omega \rightarrow \omega_0  \pm \frac{\mu_B g_J B_0}{\hbar}$, where $\mu_B=\frac{|e| \hbar}{2m_e}$ is the Bohr magnetion, $e$ is the electron charge, $m_e$ is the electron mass, and $g_J$ is the Land\'{e} $g$-factor for a given transition. When the Zeeman shift is small compared to the linewidth, we can approximate 
\begin{equation}
\label{eqderivapprox}
n_+(\omega) - n_-(\omega) \approx \frac{\mathrm{d} n(\omega)}{\mathrm{d}\omega} \; \left(\frac{2 \mu_B g_J B_0}{\hbar}\right).
\end{equation}
In Sec.~\ref{errors}, we explore the differences between the derivative approximation and the (exact) difference forms of the resonance lineshape in order to assess potential systematics associated with the lineshape model. According to Eq.~(\ref{eqderivapprox}), the Faraday rotation lineshape follows a symmetric derivative-of-dispersion shape. Its amplitude is also proportional to the atomic density, $N$, and the appropriate quantum mechanical linestrength factor, $\langle T \rangle^2$,
\begin{equation}
\label{eq3}
n(\omega) \propto N \langle T \rangle^2 \;  \frac{2 \mu_B g_J B_0}{\hbar} \; \frac{\text{d}}{\text{d}\omega}\left( \frac{ \omega - \omega_0}{(\omega - \omega_0)^2 + \Gamma^2/4}\right),
\end{equation}
where $\Gamma$ is the homogeneous linewidth (due here to collisional broadening). Finally, we must convolve this function with a normalized Gaussian, accounting for the velocity distribution of the atomic ensemble. We define the convolved lineshape, $\mathcal{L}$, as follows:
\begin{align}
\label{eqL}
\mathcal{L} (\omega,\omega_0, \Gamma, \sigma) &\equiv \frac{\mathcal{C}}{\sigma\sqrt{2\pi}} \int_{-\infty}^{+\infty} \frac{\text{d}}{\text{d}\omega}\left( \frac{ \omega - \omega^\prime}{(\omega - \omega^\prime)^2 + \Gamma^2/4}\right) \nonumber \\ 
&\times \exp{\left[\frac{-(\omega^\prime - \omega_0)^2}{2\sigma^2}\right]}  d\omega^\prime,
\end{align}
where $\sigma$, the Doppler width, is proportional to the laser frequency and the root-mean-square velocity of the hot atoms. Our experimental optical rotation spectra are carefully calibrated in terms of radians. We fit our spectra to a lineshape of the form of Eq.~(\ref{eqL}) (see Sec.~\ref{results}) allowing the amplitude scaling factor $\mathcal{C}$ to link the numerical value of the integrand on resonance (which itself is a function of the component widths) to the peak value of the experimental spectrum. Making use of Eqs.~(\ref{eq1}--\ref{eq3}) and ignoring a number of common numerical factors and fundamental constants, we find the following expression for the ratio of Faraday rotation amplitude factors:
\begin{equation}
\label{Eqratio}
\frac{\mathcal{C}_{\mathrm E2}}{\mathcal{C}_{\mathrm M1}} = \frac{\omega_{\mathrm E2}}{\omega_{\mathrm M1}} \frac{(B_0 \ell N)_{\mathrm E2} \; g_J^{\mathrm{E2}}}{(B_0 \ell N)_{\mathrm{M1}} \; g_J^{\mathrm M1}}
\frac{\lvert \langle ^3\!P_2,m=1 | E2 | ^3\!P_0 \rangle \rvert^2}{\lvert \langle ^3\!P_1,m=1 | M1 | ^3\!P_0 \rangle \rvert^2}.
\end{equation}

Here $\omega_{\mathrm E2}$ ($\omega_{\mathrm M1}$) is the resonant frequency for the \SI{939}{\nano\meter} (\SI{1279}{\nano\meter}) transition.

To find the matrix elements in~\eref{Eqratio} for many-electron states, we define the electric quadrupole and magnetic dipole moment operators, $Q_\nu$ and ${\bm \mu}$ as the sum of one-particle operators,
\begin{eqnarray}
Q_{\nu} &=& -|e| \sum_{i=1}^{N_e} \left[ r_i^2\,C_{2\nu}({\bf n}_i) \right], \nonumber \\
{\bm \mu} &=& -\frac{\mu_B}{c}  \sum_{i=1}^{N_e} [{\bf j}_i + {\bf s}_i] ,
\label{eq:Qdef}
\end{eqnarray}
where $N_e$ is the number of the electrons in the atom, ${\bf n}_i \equiv {\bf r}_i/r_i$, and $r_i$ is the radial position of the $i$th electron. ${\bf j}_i$ and ${\bf s}_i$ are the unitless total angular momentum and spin of the $i$th electron, as defined in~\cite{Sobelman1979}, and $C_{2\nu}({\bf n}_i)$ are the normalized spherical harmonics~\cite{Varshalovich}. While the sums in Eq.~(\ref{eq:Qdef}) extend over all electrons, in practice the valence $p$ electrons provide the main contribution to the matrix elements for the case of Pb.

Though, in general, each amplitude factor is proportional to the interaction length, we work hard to ensure that both laser beams traverse nearly identical physical paths through the cell. We also alternate scans in a sequence that minimizes drift-related systematic errors associated with density and magnetic field changes (see Sec.~\ref{results}). We have inserted into Eq.~(\ref{Eqratio}) matrix elements for the $E2$ and $M1$ transitions that reflect the $|\Delta m| = 1$ selection rule appropriate to the transitions we study. We make use of the fact that the matrix elements are the same for $\Delta m = +1$ and $\Delta m = -1$ for both the $E2$ and $M1$ transitions. It is possible, when the laser beam propagation direction is not precisely collinear with the B-field axis, for the $E2$ transition to exhibit small $\Delta m = \pm 2$ components, and potential consequences of this are discussed below in Sec.~\ref{errors}.

Assuming then that the relevant path length, atomic density, and magnetic field are identical for sequential laser scans for the two transitions, so that $(B_0 \ell N)_{\rm E2} = (B_0 \ell N)_{\rm M1}$, we arrive at an expression for the (unitless) quantum mechanical transition amplitude ratio, $\chi$, in terms of experimental amplitudes, resonant frequencies, and $g$-factors:
\begin{equation}
\chi \equiv \left| \frac{ \langle ^3\!P_2,m=1 | E2\, | ^3\!P_0 \rangle}{\langle ^3\!P_1,m=1 | M1\, | ^3\!P_0 \rangle} \right|=
\sqrt{\frac{\mathcal{C}_{\mathrm{E2}} \; \omega_{\mathrm{M1}} \; g_J^{\mathrm{M1}}  }{\mathcal{C}_{\mathrm{M1}} \; \omega_{\mathrm{E2}} \; g_J^{\mathrm{E2}}}}.
\label{Eqfinal}
\end{equation}
A comparison of this expression with the theory prediction will be presented below in Sec.~\ref{discussion}.

The $g$-factors are well known \cite{Porsev2016}, so that the statistical uncertainty in our ratio, $\chi$, is entirely determined by the results of our lineshape fits which determine $\mathcal{C}_{\mathrm{E2}}$ and $\mathcal{C}_{\mathrm{M1}}$.

\section{Experimental Details}\label{experiment}
\subsection{Furnace and Vapor Cell}\label{furnace}

\begin{figure}
\includegraphics[width=\columnwidth]{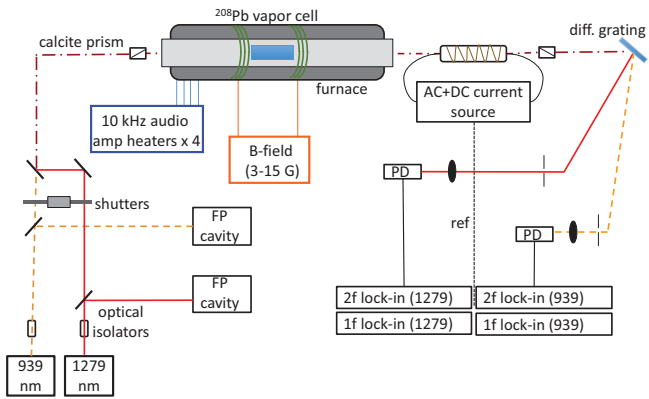}
\caption{Schematic of the experimental setup. Two commercial external cavity diode lasers (ECDLs) are scanned across the transitions' center frequencies in a sequence determined by computer-controlled shutters. The laser scans are monitored using a pair of Fabry-P\'{e}rot cavities. A calcite prism linearly polarizes the light before the furnace, after which the polarization is modulated and analyzed using a second calcite prism. The transmitted light is separated using a diffraction grating prior to detection. See text for further details.}
\label{fig_schematic}
\end{figure}

A schematic of the experimental layout is shown in Fig. \ref{fig_schematic}. The centerpiece of the experiment is the furnace, in the middle of which sits a 1-inch-diameter, 6-inch-long evacuated quartz vapor cell, containing a small quantity of isotopically enriched $^{208}$Pb (99.9\% purity). The quartz cell windows are welded to the body at \SI{10}{\degree} angles to eliminate the possibility of etalon effects in the optical path. Because of the inherent low vapor pressure of lead and the weak transition amplitudes being studied, we focus on temperatures in the 800--940\si{\celsius} range where the density is sufficiently high for easily detectable optical rotation signals. This is achieved using four ceramic clamshell heaters, which surround a meter-long ceramic tube that contains the cell. The tube is sealed at both ends with endcaps that include fused silica windows, and is evacuated and backfilled with \SI{20}{\torr} of argon in order to minimize optical beampath fluctuations due to convective air currents. A function generator operating at 10~kHz drives four audio amplifiers, which in turn drive the heaters. The frequency is sufficiently high that it does not interfere with the lock-in detection and signal analysis described below. Two thermocouple probes are positioned at the center of the vapor cell and one of the edges, which provide a temperature estimate, as well as a measure of temperature uniformity. A software \emph{p-i-d} servo loop controls the amplitude of the function generator signal, allowing us to set and stabilize the oven temperature. The furnace contains a pair of Helmholtz coils to apply the magnetic field used to create the Faraday rotation signal, and the entire assemply is enclosed in $\mu$-metal magnetic shielding. The roughly 100-fold reduction in ambient field afforded by the shielding is sufficient to bring magnetic field fluctuations to a negligible level, especially since we take the difference between sequential magnetic field-on and field-off laser scans.

\subsection{Optical Setup}\label{setup}

Two commercial external cavity diode lasers (ECDLs) at \SI{939}{\nano\meter} ($E2$) and \SI{1279}{\nano\meter} ($M1$) (Toptica DL pro series and Sacher Lasertechnik Lynx series, respectively) pass through optical isolators before a small fraction of each is directed into one of two Fabry-P\'{e}rot (FP) cavities which monitor the frequency scan range and linearity. The confocal FP cavities (finesse near 30) are constructed with invar spacers, and contained inside insulated boxes for passive thermal stabilization. The cavity free spectral ranges for the $E2$ and $M1$ lasers were independently calibrated and measured to be \SI{361.0(2)}{\mega\hertz} and \SI{501.0(3)}{\mega\hertz}, respectively. A pair of shutters allow measurements of the two transition to be made in quick succession. The beam paths are combined using a dichroic filter, and directed first through a calcite prism polarizer, then into the furnace and through the vapor cell interaction region.

Upon exiting the furnace, the laser beams pass through a 1-cm-diameter, 5-cm-long glass rod with a large Verdet constant (``Faraday glass'') which is contained within a solenoid to which we can apply AC and DC currents, thus either modulating or tilting the laser polarization. We typically drive the solenoid with \SI{2}{\ampere} of AC current at $\omega = 2\pi \times 500~$\si{\hertz}, which results in a polarization modulation amplitude of a few milliradians. The laser beams then traverse a second, crossed calcite polarizer. Our polarizer pair in isolation has a finite extinction ratio of better than $10^{-6}$, but the presence of the furnace, vapor cell windows, and Faraday glass limit the effective extinction ratio of our polarimeter to about $2\times 10^{-5}$. The polarizers are each housed in a rotational lever mount actuated with a differential micrometer. Given the geometry of our mount and the \SI{1}{\micro\meter} resolution of the differential micrometer, we can reliably set and control the polarizer tilt angle at the \SI{10}{\micro\radian} level. 

The light is then incident on a diffraction grating, which separates the two laser beam paths. With the aid of collimators and lenses, we focus each laser beam onto a high-gain, low-noise photodiode detector. This arrangement also allows us to reject nearly all of the substantial (but incoherent) blackbody radiation emanating from the furnace. This is important given that the coherent laser radiation reaching our detector after exiting the polarimeter is never more than about \SI{100}{\nano\watt}.

\subsection{Modulation, Lock-in Detection, and Calibration}\label{detection}

The detection scheme, similar to that described in \cite{Meekhof1995}, uses the modulator combined with a pair of lock-in amplifiers for each wavelength in order to extract the optical rotation signal. After passing through the atomic vapor, the laser intensity is $I(f)$, reflecting the absorption lineshape, and there is also a frequency-dependent rotation of $\Phi_{\text F}(f)$, due to the atomic Faraday effect. We also account for a small frequency-dependent optical birefringence, $\Phi_{\text{br}}(f)$, unrelated to the atoms. The Faraday modulator introduces an additional sinusoidal rotation of $\Phi_{\mathrm{rot}} \cos(\omega t)$. The resulting intensity through the second polarizer is thus (using the small angle approximation):

\begin{align}
I_{\text{out}} &= I(f) \sin^2 \left[ \Phi_{\text{Pb}}(f) + \Phi_{\text{br}}(f) + \Phi_{\text{rot}} \cos(\omega t) \right] \nonumber\\
&\approx I(f) \left[ \Phi_{\text{Pb}}^2(f) + \Phi_{\text{br}}^2(f) + 2\Phi_{\text{Pb}}(f)\Phi_{\text{br}}(f) \right. \\
&+ \left. 2 \Phi_{\text{rot}} \cos(\omega t)(\Phi_{\text{Pb}}(f) + \Phi_{\text{br}}(f)) + \Phi^2_{\text{rot}}\cos^2(\omega t) \right], \nonumber
\end{align}
where we have ignored the small constant transmission component from the finite polarimeter extinction. This expansion results in three important components: a constant term, one oscillating at $\omega$, and another oscillating at $2\omega$. Lock-in detection at $\omega$ and $2\omega$ removes the DC term; the $2\omega$ term is only dependent on the transmitted intensity, whereas the $1\omega$ term is proportional to the Faraday optical rotation times the transmission. Thus, the ratio of the two signals $S_{1\omega}/S_{2\omega}$ yields a signal proportional to the optical rotation only. Four lock-in amplifiers (Stanford Research Systems SRS 810) are set to the fundamental and second harmonic of the modulation frequency for the two lasers, and the extracted signals from the four are collected using a data acquisition board.

The size of the lock-in signal we detect is also proportional to the the amplitude of the modulation, $\Phi_{\mathrm{rot}}$. However, we know that the Verdet constant of our Faraday glass is substantially different at our two laser frequencies. To account for this in our calibration procedure, we first perform the following off-line exercise for each laser in turn. We fix the laser frequency at a value away from the atomic resonance.  While still modulating the magnetic field, we add a stepwise series of increasing DC currents to the solenoid. At each step, we use the micrometer controlling the second polarizer to `re-cross' the polarimeter by noting when the $1\omega$ lock-in output reaches exactly zero. In this way, we can accurately find the ratio of the rotatory effects of the Faraday glass for our two laser frequencies. Repeated calibration exercises such as these were performed over the one-month period of data collection to study reproducibility upon laser beam and polarimeter realignment. With these measurements in hand, we can, as noted below, incorporate a second procedure into our data collection sequence in which we apply a large, discrete DC current step to the solenoid, and, while directing both lasers through the cell (at fixed frequencies), detect the corresponding step-size changes in the lock-in outputs. When we include the results of both calibration procedures, we can then convert the units of the experimental signal of interest (ratio of lock-in outputs) to absolute radians for each transition.

\subsection{Data Acquisition Procedure}\label{acquisition}

Data acquisition was performed at a range of temperatures (\SI{800}{\celsius} -- \SI{940}{\celsius}) and with a range of applied currents to the Helmholtz coils (\SI{1}{\ampere} -- \SI{4}{\ampere}). Acquisition was done in three steps: an initial calibration sequence, the main measurement sequence, and a final calibration sequence. The main measurement sequence has eight components and is typically looped five times. Table \ref{sequence} summarizes the data collection sequence. We refer to this as a `run.' The goal of the sequence is to examine possible sources of systematic error by measuring each transition's rotation with a background scan without an applied magnetic field either immediately before or immediately after the field is applied and the rotation is measured. We acquire field-on / field-off scans, and also $E2$ and $M1$ scans in an ``ABBA'' sequence configuration to allow us to study and minimize temporal drift-related systematic errors. Such a collection sequence typically required one hour to complete.

\begin{table}[htbp]
	\centering
		\begin{tabular}{|c|c||c|c|c|c|c|c|c|c||c|}
		\hline
		\textbf{Data} & Cal. & \multicolumn{8}{c||}{Frequency Scan $(\times 5)$} & Cal. \\
		\hline
		$\mathbf{\lambda}$ & $E2$/$M1$ & $E2$ & $E2$ & $M1$ & $M1$ & $M1$ & $E2$ & $E2$ & $M1$ & $E2$/$M1$ \\
		\hline
		$\mathbf{B_{\text ext}}$ & & & x & x & & x & x & & & \\
		\hline
		\end{tabular}
	\caption{Data acquisition sequence. Each individual up/down scan pair takes 15 seconds. The `x' in the $\mathbf{B_{\text ext}}$ row reflects application of the longitudinal magnetic field to the atoms.}
	\label{sequence}
\end{table}

An individual scan is based upon a triangle wave applied to a laser's intracavity piezoelectric transducer (PZT), changing its frequency and scanning across the transition's linecenter, which typically requires \SI{20}{\second} to complete. The atomic spectral features of interest extend over roughly \SI{1}{\giga\hertz}, and a typical laser scan extended over \SI{4}{\giga\hertz}. We separately analyzed the frequency-increasing portion of the scan (``upscan'') as well as the portion with a downward slope (``downscan''). For each run with a particular laser, a data acquisition computer recorded the triangle voltage wave, the transmission of the Fabry-P\'{e}rot cavity the $1\omega$ lock-in amplifier signal, and the $2\omega$ lock-in amplifier signal.

At each temperature, we acquired between 4 and 6 runs, between which optical realignments, changes of laser beam powers, and changes in laser sweep characteristics were applied. In all, roughly \SI{40}{\hour} of data were collected, representing 800 distinct $E2$/$M1$ amplitude ratio measurements. The temperature range over which we worked corresponds to more than an order of magnitude change in lead vapor density. The corresponding $M1$ Faraday rotation amplitudes range from \SI{2}{\milli\radian} to \SI{50}{\milli\radian}, while the $E2$ amplitudes were in the \SI{200}{\micro\radian} to \SI{5}{\milli\radian} range.

\section{Data Analysis and Results}\label{results}
\subsection{Data Analysis Procedure}\label{analysis}

The first step in data analysis involves using the Fabry-P\'{e}rot transmission data to linearize and calibrate the frequency scans. Using the FP peak locations, we model the frequency as a fourth-order polynomial function of scan point number to account for small nonlinearity in the PZT voltage response. We found that higher-order polynomials did not improve the statistical quality of the FP peak fits. Using this frequency axis, we construct the unitless ratio of the $1\omega$ to $2\omega$ lock-in outputs, and then apply the calibration factors described above to convert this ratio to units of radians. In each case, we use the average step calibration values obtained by the pre- and post-calibration scans for that particular data run. This procedure is applied to both the $M1$ and $E2$ scans for both the field-on and field-off configurations. We next subtract the field-off scans proximate to the associated field-on scan, removing background features unrelated to the atoms that are typically a few percent of the field-on Faraday signals.

\begin{figure}
\includegraphics[width=.9\columnwidth]{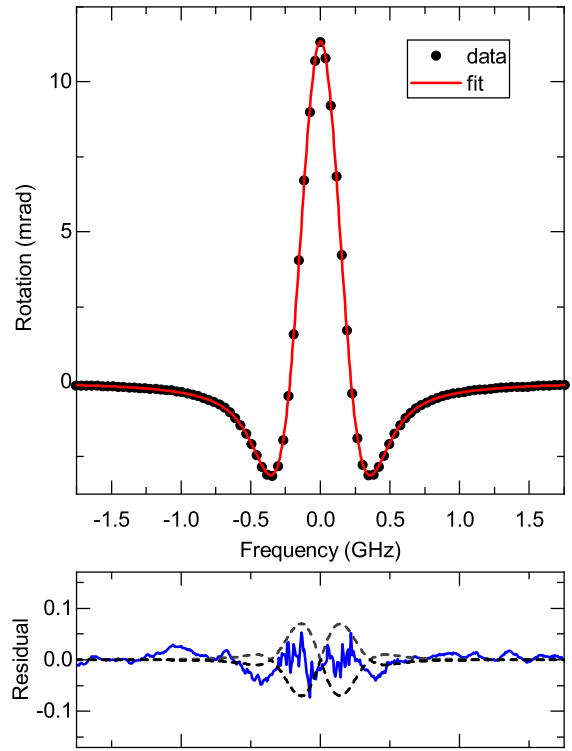}
\caption{Sample data from \SI{800}{\celsius} $M1$ Faraday rotation signal (black dots, every fifth point shown) and fit result (red line). Residuals, expanded by a factor of 20, are shown below; solid blue shows the unweighted residual, while the dashed black line shows the envelope of the noise expected from a model that includes laser frequency jitter (see text).}
\label{fig:M1_sample}
\end{figure}

\begin{figure}
\includegraphics[width=.9\columnwidth]{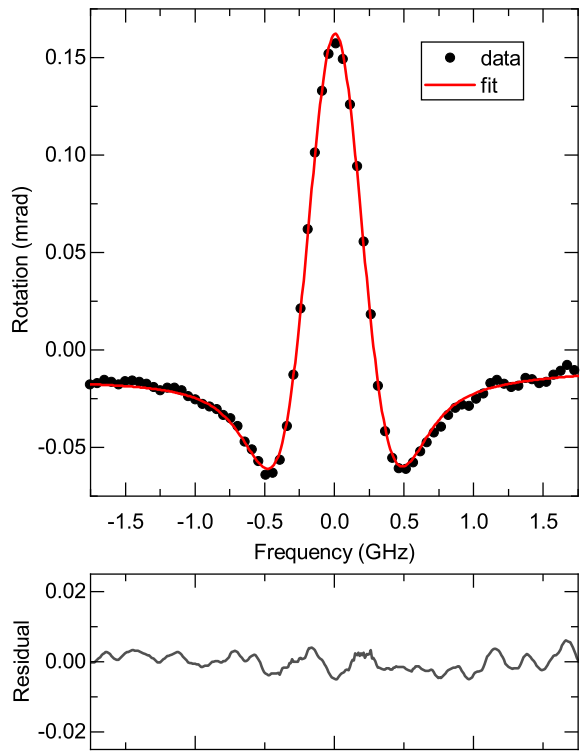}
\caption{Sample data from \SI{800}{\celsius} $E2$ Faraday rotation signal (black dots, every fifth point shown) and fit result (red line). Expanded residuals are shown below.}
\label{fig:E2_sample}
\end{figure}

The subtracted lineshape is then fitted using a standard nonlinear least squares algorithm to the convolution function described in Eq.~(\ref{eqL}). With two thermocouple temperature monitors near the cell, we have a fairly accurate estimate of the temperature. We choose, then, to fix the Doppler width to a calculated value for the case of each laser scan. Below we discuss our exploration of lineshape changes and associated systematic amplitude errors resulting from our estimated temperature uncertainty. We note that, since ultimately we determine the ratio of the $E2$ to $M1$ amplitudes, overall temperature uncertainty largely cancels in this ratio, since the ratio of Doppler widths is temperature-independent. We therefore analyzed our Faraday lineshapes by fitting to two key parameters: the Lorentz width, $\Gamma$, due here to lead-lead collisional broadening, and the amplitude parameter $\mathcal{C}$ introduced in Sec. \ref{background}, connecting our convolution lineshape to the experimental peak height. We find this homogeneous linewidth component to be roughly ten times smaller than the Doppler width for the case of both transitions. In order to account for imperfect background subtraction, we also add constant and linear background parameters to the fit, which are always quite small, and, in the case of the linear term, often statistically unresolved. Examples of single background-subtracted scans of each transition at \SI{800}{\celsius} (near the low end of our temperature range) are shown in Figs. \ref{fig:M1_sample} and \ref{fig:E2_sample}, along with the residuals of the fits. Each scan shown represents about \SI{40}{\second} of data collection. As one can see, the residual RMS optical rotation noise is at the few \si{\micro\radian} level in both cases. Because of its much larger amplitude, the $M1$ scan exhibits a baseline signal-to-noise ratio of more than 1000:1. Interestingly, in this case there is a significant increase in the size of the residuals near linecenter. In fact, this can be easily modeled as an effective amplitude noise induced by short-term frequency jitter of the diode laser as it scans across the transition --- something that would manifest in the regions of the lineshape where the slope is steepest. The dashed envelope included in the lower box of Fig, \ref{fig:M1_sample} shows the expected amplitude noise from a frequency jitter of \SI{1}{\mega\hertz} --- something quite typical of ECDL systems such as ours.

\begin{figure}
\includegraphics[width=.9\columnwidth]{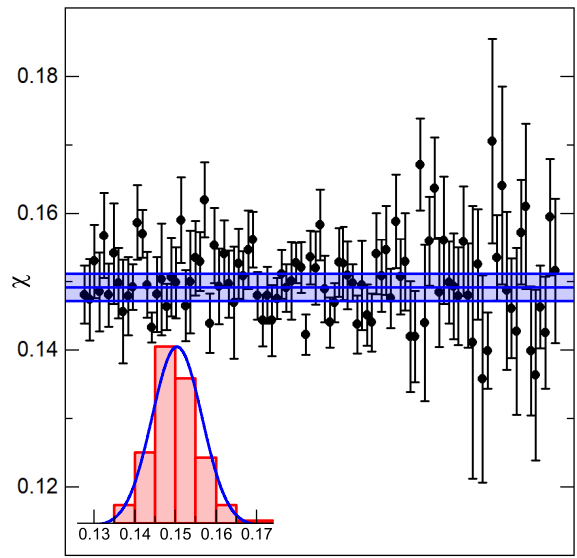}
\caption{Distribution of $\chi$ measurements from all 95 individual scan ratios taken at \SI{800}{\celsius}. Main figure: $\chi$ and corresponding error bars, with mean and standard deviation (solid blue) shown. Inset: histogram of $\chi$ (red bar plot) and a fitted Gaussian (thick blue curve). Intrinsic precision of $\chi$ values varies for subsets of these data depending, for example, on magnetic field employed for a given run.}
\label{fig:histogram}
\end{figure}

\begin{figure}
\includegraphics[width=.9\columnwidth]{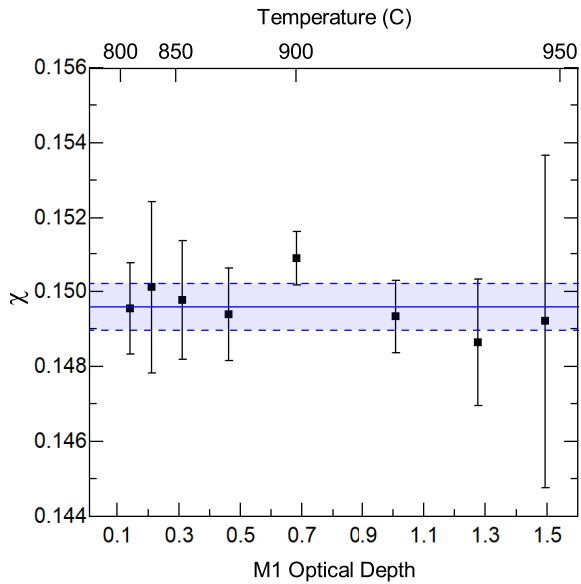}
\caption{The amplitude ratio $\chi$ as a function of optical depth (bottom axis) and temperature (top axis). The mean and standard error are shown in solid and dashed blue lines, respectively.}
\label{fig:ratio}
\end{figure}

Fit results are organized by laser scan direction and order of field-on / field-off sequencing. We scale our amplitude fit parameters using the calibration factors discussed above. The difference between the pre- and post-calibration scans within a data run yields a measure of calibration uncertainty, which can be combined with the error bar generated by the fit procedure to arrive at a final uncertainty for the corrected fit amplitude. We then construct the ratio of the fit amplitudes for the two transitions, $\mathcal{C}_{\mathrm{E2}}/\mathcal{C}_{\mathrm{M1}}$, for each set of consecutive $E2$ and $M1$ scans. Inserting the values for the ratios of the frequencies and $g$-factors of these transitions, we finally obtain experimental values for $\chi$ as defined in Eq.~(\ref{Eqfinal}). We accumulated statistics on all amplitude ratios taken at a given temperature. In some cases, the scatter between the weighted mean value for different data runs at a given temperature slightly exceeded their respective standard errors, due, for example, to small changes in experimental conditions, thermal drift, or relative beam path changes of the two lasers due to purposeful optical realignment. In each case, we expanded our error bars to account for this measured variance. We also took the approach of generating a histogram for all values at a given temperature and fitting this distribution to a Gaussian (see Fig. \ref{fig:histogram}). The mean values arrived at by these two methods agreed very well within statistical uncertainties.  Fig. \ref{fig:ratio} shows the complete data set for our measured values of $\chi$ plotted as function of temperature, and corresponding $M1$ absorptive optical depth. Final weighted mean and $1\sigma$ statistical uncertainty are indicated in blue. This range of temperatures corresponds to a roughly a factor of 15 in vapor density, and as such is associated with changes in amplitude, and component spectral widths of the Faraday lineshape. Below \SI{800}{\celsius}, the $E2$ amplitudes were too small to achieve reliable fit results. At the upper end of our temperature range, where relatively large rotation amplitudes should have provided the best statistical precision, we observed a large scan-to-scan and run-to-run variation in ratios, likely due to much larger thermal drifts and optical birefringence effects at this temperature. Ultimately, while the mean value at \SI{940}{\celsius} agrees well with other data sets, the uncertainty is significantly higher due to this increased scatter, and we did not seek to increase the temperature further.  Our final value and $1\sigma$ statistical uncertainty in the measured ratio is $\chi =0.1496(7)_{\text{stat}}$.

\subsection{Exploration of Systematic Errors}\label{errors}

Potential systematic errors in our experimental value for $\chi$ were studied extensively, and the results are summarized in Table \ref{error_table}, where systematic error contributions to the unitless ratio are expressed in percentages. Potential error sources which did not show statistically resolved effects are listed with a dash. In many cases, the fact that we are taking the ratio of two amplitudes tends to reduce systematic error impact (such as for temperature uncertainties, or magnetic field inhomogeneities). In addition, since $\chi$ is proportional to the square root of the amplitude ratio, the size of potential errors in $\chi$ associated with extracting Faraday signal amplitude are immediately reduced by a factor of two. Further, $1/f$-type noise associated with thermal, mechanical, or magnetic field drifts occurring on time scales comparable to our scan sequence would tend to show up as increased scatter between measurements rather than systematic bias, especially given our choice of sequencing repeated measurements in an ``ABBA'' pattern.

As can be seen in Fig. \ref{fig:ratio}, we importantly do not see a resolved systematic trend in our measured ratio as a function of temperature/density. In addition to comparing results at a number of different temperatures, we compared results for laser scan direction, different scan speeds and scan ranges, different temporal order of field-on/field-off scans, and different temporal order of $E2$ vs. $M1$ scans. Occasionally, we saw comparisons of subsets of data that differed by 1.5 to 2.0$\sigma$, where $\sigma$ is the combined error of the data subsets, and these contributions to the net systematic uncertainty are included in the table.  

\begin{table}[htb]

\vspace{10pt}
\begin{tabular}{|l|c|}
\hline
\textbf{Source} & \textbf{Error in $\chi$ (\%)} \\
\hline
\textbf{Statistical error} & 0.48 \\
\hline
\textbf{Fitting} & \\
Frequency Linearization & 0.02 \\
Fixing vs. Floating Lorentz Widths & 0.37 \\
Linear Background & 0.27 \\
Lineshape Weighting & 0.32 \\
Incorrect Doppler Widths & 0.10 \\
Include / Discount Scan Wings & 0.13 \\
\hline
\textbf{Signal Modeling} & \\
Derivative vs. Difference & 0.19\\
Magnetic Field Dependence & -- \\
\hline
\textbf{Geometry} & \\
$E2$ $\Delta m = 2$ Transitions & 0.35 \\
\hline
\textbf{Laser Scanning Properties} & \\
Scan Direction  & 0.22\\
Scan Speed / Width & -- \\
\hline
\textbf{Data Collection} & \\
Field-On / Field Off Order & --\\
$E2$ / $M1$ Order & 0.13 \\
\hline
\textbf{Angle Calibration} & \\
Off-Line Calibration & 0.28\\
Pre/Post Variance & 0.29 \\
\hline
\textbf{Other} & \\
Isotopic Purity & 0.02 \\
\hline
\textbf{TOTAL:} & \textbf{0.98\%} \\
\hline
\end{tabular}
\caption{Summary of error contributions and sources, expressed as percentage errors in the experimental ratio $\chi$. Horizontal line entries reflect the lack of a resolved systematic error contribution.}
\label{error_table}
\end{table}

\subsubsection{Calibration}
Errors in any aspect of our calibration procedure would directly impact our amplitude ratio measurement, and these were explored as follows. We completed several of the off-line calibration exercises over the course of our data-collection period, and compared the ratio of calibration factors obtained in these procedures. We assign a systematic error component based on the variation of these measurements (most likely due to small changes in the relative optical paths of the lasers, or possibly small thermal drifts over the time scale of the measurements). Also, for all of our data collection runs, we studied the differences between the pre- and post-calibration scans to estimate the potential errors associated with using their average to calibrate all scans in that run. An estimate of the systematic error associated with taking our approach of calibrating all runs based on the average of the two calibration values is also included in Table \ref{error_table}.

\subsubsection{Fitting Methods}
We explored a number of alternative approaches to fitting our Faraday rotation spectra to quantify systematic effects associated with lineshape analysis. First, as noted above, we explored different polynomial orders for parametrization of the ECDL scan nonlinearity, finding that beyond fourth order, no statistically significant changes to the fitted amplitude were seen. Our nominal method for fitting our spectra involved equal weighting of all points in the scan. We explored two alternatives. First, we explored a model that weighted data points according to a model that accounted for the frequency noise and associated fluctuations as noted in Fig. \ref{fig:M1_sample} and discussed in the previous section. Second, we explored truncating our fit ranges to exclude portions of the scan farther away from the resonant lineshape. We saw small changes in our fitted amplitude results, always well below the 1\% level, and include small error contributions for these in Table \ref{error_table}. 

We studied the reliability of our Lorentzian and Gaussian (Doppler) width determinations in detail. Possible errors in these parameters impact the peak value of the lineshape convolution function defined in Eq.~(\ref{eqL}), and thus directly affect the fitted amplitude parameters, $\mathcal{C}$, from which we determine $\chi$. As noted, since the ratio of the Doppler widths for the two transitions is temperature-independent, a potential systematic error in $\chi$ due to temperature error could only come from the secondary effect of producing associated changes in other fit parameters that would affect the two transitions lineshapes in different ways. We explored this by systematically choosing a temperature (and hence Doppler widths) over a $\pm$20-degree range centered on the nominal temperature (which is taken to be the average of our two thermocouple readings). We then fit both experimental lineshapes, extracting the Lorentzian width, $\Gamma$, and peak amplitude factor, $\mathcal{C}$, in our usual fashion. Even using this relatively large temperature range, roughly equal to the difference in our thermocouple readings, we saw changes in the value of $\chi$ only at the $\pm 0.1\%$ level, and have included this in our error table.  

\begin{figure}
\includegraphics[width=.9\columnwidth]{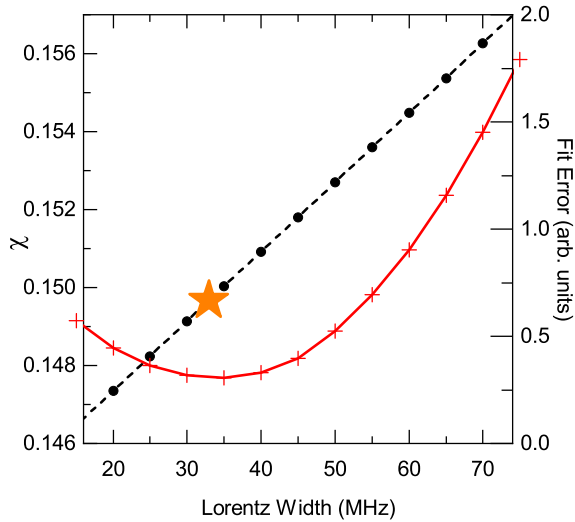}
\caption{Exploration of a potential systematic error from a fixed, miscalculated Lorentzian width. The amplitude ratio, $\chi$, is plotted with black dots and dashes on the left y-axis. The corresponding fit error for those values and Lorentz widths are plotting with red plusses, on the right y-axis. The orange star indicates the Lorentz width and $\chi$ of a floated Lorentz width. Further details are provided in the text.}
\label{fig:gamma_sys}
\end{figure}

Of more concern is the accuracy of our Lorentz width determinations. These widths are an order of magnitude smaller than the Doppler widths, and thus more challenging to extract. However, their value clearly affects the amplitude of our lineshape function (Eq.~(\ref{eqL})). Our standard analysis method starts with fixed Doppler widths and optimizes the Lorentz width parameter in the fit process. In order to explore the effect of potential errors in Lorentz width values on our ratio $\chi$, we proceeded as follows. Since the $E2$ Faraday amplitudes have substantially lower signal-to-noise ratio, we assumed, for the purpose of this exercise, that the standard $M1$ fit procedure is able to extract the `correct' Lorentz width, $\Gamma_{\mathrm{M1}}$. Then, we fit $E2$ lineshapes using a modified procedure where instead we fix the Lorentz width (in addition to the Doppler width) to a series of values above and below the apparent `best fit' value, and allow only the peak amplitude factor to be optimized (this optimized value is clearly correlated with the choice of $\Gamma_{\mathrm{E2}}$). We then recorded the summed chi-squared value of the overall lineshape fit for each fixed choice of $\Gamma_{\mathrm{E2}}$. Figure \ref{fig:gamma_sys} summarizes this exploration for the case of all the data runs taken at one temperature (here \SI{900}{\celsius}). The red curve indicates the changing `quality of fit' for the entire collection of fits at \SI{900}{\celsius} at each fixed choice of $\Gamma_{\mathrm{E2}}$. The black line simply maps out the correlation between $\chi$ and $\Gamma_{\mathrm{E2}}$ , assuming that $\Gamma_{\mathrm{M1}}$ remains constant. The orange `star' shows the average Lorentz width parameter generated by our standard fitting procedure, in which $\Gamma_{\mathrm{E2}}$ is `floated.' The excellent agreement between the two methods in terms of finding the optimal value $\Gamma_{\mathrm{E2}} \approx$ \SI{35}{\mega\hertz} is reassuring, and we can see even a very large fractional change in $\Gamma_{\mathrm{E2}}$ of $\pm$ \SI{10}{\mega\hertz} that yields a change in $\chi$ of only $\pm 1\%$. A more extensive analysis of data sets at all temperatures allows us to place a $\pm 0.4\%$ systematic error based on our estimated uncertainty in the the extracted Lorentz widths.

\subsubsection{Lineshape Model}
We also considered the systematic error associated with using the derivative approximation to the Faraday lineshape. First, for a series of Zeeman splittings in our experimental range, we generated theoretical lineshapes with typical values for component widths using the difference (rather than the derivative) of the dispersive real part of the refractive index lineshapes. We then proceeded to fit these lineshapes using our standard (derivative approximation) fitting function and studied the changes in fitted amplitude as a function of the Zeeman splitting. Since the $g$-factors and component widths of the two transitions are different, this would impact the two transitions differently, and hence would produce a systematic error in $\chi$. From this investigation, we put a limit of the potential systematic error of our derivative approximation at the 0.2\% level. As a second experimental check, we studied the correlation of $\chi$ with the current applied to the Helmholtz coils for the data we collected. This showed no statistically resolved trend over the $\approx$~3--\SI{15}{\gauss} range of magnetic fields that we explored.

\subsubsection{Geometrical Misalignment}
Finally, we note that our analysis assumes that the laser beam paths are exactly collinear with the magnetic field axis within the vapor cell interaction region. This effectively allows us to view the electric quadrupole interaction as an operator proportional to the $\ell = 2, m = 1$ spherical harmonic (see Sec. \ref{discussion} below). For small deviations from collinearity, $\delta\theta$, one can show that $\Delta m = \pm 2$ transitions are possible, and that the size of these components relative to the dominant $\Delta m=\pm 1$ transitions is proportional to $|\sin(\delta\theta)|$ \cite{Roos2000}. Given our apparatus geometry and laser beam collimation, we estimate that $|\sin(\delta\theta)| \le 2^{\circ}$. We were able to explore the implications of this by generating simulated Faraday rotation spectra with small $\Delta m = \pm2$ components, and then analyzing these modified lineshapes using our standard fitting routine. By studying the impact of this non-ideal geometry on the fitted lineshape amplitudes, we can place a limit on its potential systematic error contribution to $\chi$, which is included in Table \ref{error_table}. We note that, even with perfect collinearity, small stray magnetic fields, either from external sources or mu-metal remanence, would ultimately produce a small systematic geometric uncertainty. For the experimental fields employed here, we estimate this contribution to misalignment to be several times smaller than the current optical collinearity contribution.

We lastly mention that such geometrical misalignment also produces more complicated magneto-optical effects, including the so-called `Voigt' effect. As discussed in detail in \cite{Edwards1995}, the size of these additional components, given the estimated size of our misalignment, would produce changes to our Faraday lineshape that are well below our level of statistical sensitivity.

\subsubsection{Isotopic Purity}
Given the quoted isotopic purity of the vapor cell (99.9\%), we generated realistic simulated lineshapes and fit these using our standard analysis procedure to produce the systematic relevant error estimate in Table \ref{error_table}.

\subsubsection{Final Experimental Ratio}
Combining all of the systematic error contributions in quadrature gives an uncertainty roughly twice that of the statistical error. Combining these leads to a final experimental value for our unitless amplitude ratio: $\chi = 0.1496 \pm 0.0015$. In Sec. \ref{discussion}, we establish the connection between this ratio and the reduced electric quadrupole matrix element, the \emph{ab initio} theoretical derivation for which we present next.

\section{Theory}\label{theory}

We evaluated the reduced matrix elements (MEs) of the $6p^2\,\,^3\!P_0 -\, 6p^2\,\,^3\!P_2$ and $6p^2\,\,^3\!P_0 -\, 6p^2\,\,^1\!D_2$ $E2$ transitions as well as the static scalar and tensor polarizabilities of the $6p^2\,\,^3\!P_1$ and $6p7s\,\,^3\!P_0^o$ states of Pb using the high-precision relativistic CI+all-order method~\cite{Safronova2009}. This method was adopted by us for calculating the PNC amplitude for the $6p^2\,\,^3\!P_0 -\, 6p^2\,\,^3\!P_1$ transition~\cite{Porsev2016}. 

We consider Pb as a four-valence atom. The basis set was constructed using a $V^{N-2}$ approximation in the framework of the Dirac-Fock-Sturm approach (see Ref.~\cite{Porsev2016} for more details). In this calculation, we use the wave functions obtained in~\cite{Porsev2016} in the CI+MBPT~\cite{Dzuba1996} and CI+all-order approximations. We carry out calculations in both approximations considering the CI+all-order results as the recommended ones. Atomic units ($\hbar=|e|=m=1$) are used throughout unless stated otherwise.

\subsection{$E2$ Transitions}
\label{theory_E2}

Using the expression for the electric quadrupole moment operator, given by Eq.~(\ref{eq:Qdef}), we obtain
for the $E2$ $6p^2\,\,^3\!P_0 -\, 6p^2\,\,^3\!P_2$ transition,
\begin{eqnarray}
|\langle ^3\!P_0 ||Q|| ^3\!P_2 \rangle| &\approx& 8.91\,\, {\rm a.u.}\,\, ({\rm CI+MBPT}), \nonumber \\
                                        &\approx& 8.86\,\, {\rm a.u.}\,\, ({\rm CI+all-order}) .
\end{eqnarray}

Inclusion of the Breit interaction correction increases the absolute value of the matrix element (ME) by 0.02 a.u.. The quantum-electrodynamic (QED) correction is negligible at the current level of calculation accuracy. The difference of the values obtained at the CI+MBPT and CI+all-order stages gives us an estimate of the uncertainty. Thus, the final recommended value is:
\begin{equation}
\label{eq:Qfinal}
 |\langle ^3\!P_2 ||Q|| ^3\!P_0 \rangle| = 8.88(5)~\text{a.u.}.
\end{equation}
We have also estimated the reduced ME of the electric quadrupole $6p^2\,\,^3\!P_0 -\, 6p^2\,\,^1\!D_2$ transition. This is an intercombination transition (the initial and final states have different total spin $S$). As a result, it is an order of magnitude smaller than $|\langle ^3\!P_2 ||Q|| ^3\!P_0 \rangle|$. We find
\begin{equation}
|\langle ^1\!D_2 ||Q|| ^3\!P_0 \rangle| \approx 0.63\,\, {\rm a.u.}.
\end{equation}

\subsection{Polarizabilities}
\label{theory_polar}

\begin{table*}
\caption{\label{tab_scal} Polarizabilities obtained using the CI+all-order approximation.
Contributions to the $6s^2 6p^2\,\,^3\!P_1$ and $6s^2 6p7s\,\,^3\!P_0^o$ scalar static polarizabilities, $\alpha_0$, of Pb (in a.u). The dominant contributions to the valence polarizability from intermediate states $|n\rangle$ are listed separately with the corresponding absolute values of electric-dipole reduced matrix elements given (in a.u.) in the column labeled ``$D$''. The theoretical and experimental~\cite{NIST} transition energies $\Delta E \equiv E(n) - E(6p^2\,^3\!P_1)$ and $\Delta E \equiv E(n) - E(6p7s\,^3\!P_0^o)$ are given (in cm$^{-1}$) in columns $\Delta E_{\rm th}$  and $\Delta E_{\rm expt}$. The remaining contributions to the valence polarizability are given in the row labeled ``Other.'' The values listed in the row labeled ``Total val.'' are obtained as the sum of all listed contributions and ``Other.''
The dominant contributions to $\alpha_0$, listed in columns $\alpha_0[\text{A}]$ and $\alpha_0[\text{B}]$, are calculated with CI+all-order+RPA matrix elements and theoretical [A] and experimental [B] energies \cite{NIST}, respectively.}
\begin{ruledtabular}
\begin{tabular}{cccrrcc} 
\multicolumn{1}{c}{State} & \multicolumn{1}{c}{$|n\rangle$} & \multicolumn{1}{c}{$\Delta E_{\rm th}$} & \multicolumn{1}{c}{$\Delta E_{\rm expt}$}
& \multicolumn{1}{c}{$D$} & \multicolumn{1}{c}{$\alpha_0[\text{A}]$} &\multicolumn{1}{c}{$\alpha_0[\text{B}]$} \\
\hline \\ [-0.3pc]
$6p^2\,\,^3\!P_1$ &  $6p7s\,\,^3\!P_0^o$  & 27207 &  27141 &  1.92  &  6.6  &  6.6  \\
                  &  $6p7s\,\,^3\!P_1^o$  & 27533 &  27468 &  1.41  &  3.5  &  3.5  \\
                  &  $6p6d\,\,^3\!F_2^o$  & 38222 &  37624 &  0.08  &  0.01 &  0.01  \\
                  &  $6p6d\,\,^3\!D_2^o$  & 39046 &  38242 &  3.45  & 14.9  & 15.2  \\
                  &  $6p6d\,\,^3\!D_1^o$  & 39110 &  38249 &  0.63  &  0.5  &  0.5  \\
                  &  $6p7s\,\,^3\!P_2^o$  & 40572 &  40370 &  0.78  &  0.7  &  0.7  \\
                  &  $6p8s\,\,^3\!P_1^o$  & 41737 &  40868 &  1.13  &  1.5  &  1.5  \\
                  &  $6p8s\,\,^3\!P_0^o$  & 42275 &  40907 &  0.65  &  0.5  &  0.5  \\
                  &  $6p7s\,\,^1\!P_1^o$  & 42670 &  41621 &  0.20  &  0.04 &  0.05  \\
                  &  Other                &       &        &        & 25.9  & 25.9  \\
                  &  Total val.           &       &        &        & 54.2  & 54.6  \\
                  &  Core + Vc            &       &        &        &  3.8  &  3.8   \\
                  &  Total                &       &        &        & 58.0  & 58.4  \\
\hline \\ [-0.5pc]
$6p7s\,\,^3\!P_0^o$ &  $6p^2\,\,^3\!P_1$  &-27207 & -27141 &  1.92  & -20   & -20  \\
                    &  $6p7p\,\,^3\!P_1$  &  7837 &   7959 &  3.99  & 298   & 293  \\
                    &  $6p7p\,\,^3\!D_1$  &  9605 &   9715 &  5.43  & 450   & 445  \\
                    &  $6p8p\,\,^3\!P_1$  & 17800 &  16361 &  0.17  &   0.2 &   0.2 \\
                    &  $6p8p\,\,^3\!D_1$  & 18336 &  16957 &  1.04  &   8.6 &   9.4 \\
                    &  Other              &       &        &        &  19   &  19   \\
                    &  Total val.         &       &        &        & 756   & 747   \\
                    &  Core +Vc           &       &        &        &   4.1 &  4.1  \\
                    &  Total              &       &        &        & 760   & 751
\end{tabular}
\end{ruledtabular}
\end{table*}

\begin{table}[htb]
\caption{\label{tab_polar} The static scalar ($\alpha_0$) and tensor ($\alpha_2$) polarizabilities
obtained in the CI+MBPT and CI+all-order approximations (in a.u.) are presented. The differences of the CI+all-order and CI+MBPT results are given (in \%) in the column labeled ``diff.'' The recommended values and their uncertainties are given in the last column.}
\begin{ruledtabular}
\begin{tabular}{cccccc}
        State       &             & CI+MBPT  & CI+all-order & diff(\%) & Recom.  \\
\hline \\ [-0.5pc]
$6p^2\,\,^3\!P_1$   & $\alpha_0$  &  58.7    &  58.0        &  1.2      &  58.0(7)     \\[0.1pc]
                    & $\alpha_2$  &  -5.8    &  -5.7        &  1.5      &  -5.7(1)     \\[0.3pc]
$6p7s\,\,^3\!P_0^o$ & $\alpha_0$  &  752     &  760         &  1.1      &  760(8)
\end{tabular}
\end{ruledtabular}
\end{table}

The scalar dynamic polarizability $\alpha(\omega)$ can be separated into three parts:
\begin{equation}
\alpha(\omega) = \alpha_v(\omega) + \alpha_c(\omega) + \alpha_{vc}(\omega),
\label{alpha}
\end{equation}
Where $alpha_v$ is the valence polarizability and $\alpha_c$ is the ionic core polarizability. A small term, $\alpha_{vc}$, is included due to the presence of the four valence electrons and possible excitation of a core electron to the occupied shell. Thus, $\alpha_{vc}$ serves to restore the Pauli principle and slightly modifies the core polarizability~\cite{Porsev2002}.

The valence part of the a.c. electric dipole polarizability of the $|\Phi_0 \rangle$ state can be written in
the following form:
\begin{eqnarray}
\alpha_v (\omega) &=& 2\, \sum_k \frac { \left( E_k-E_0 \right)
|\langle \Phi_0 |D_0| \Phi_k \rangle|^2 }
      { \left( E_k-E_0 \right)^2 - \omega^2 }  \nonumber \\
 &=& \sum_k \left[ \frac {|\langle \Phi_0 |D_0| \Phi_k \rangle|^2 } {E_k - E_0 + \omega}
 +\frac {|\langle \Phi_0 |D_0| \Phi_k \rangle|^2 } {E_k - E_0 - \omega} \right]\!,
\label{Eqn_alpha}
\end{eqnarray}
where $D_0$ is the $z$-component of the effective electric dipole operator ${\bf D}$, defined (in a.u.) as ${\bf D} = -{\bf r}$.
By the effective (or ``dressed'') electric dipole operator, we mean that the operator also includes the random-phase approximation (RPA) corrections~\cite{Dzuba1998}.

To account for intermediate high-lying discrete states and the continuum, we calculated $\alpha_v(\omega)$
by solving the inhomogeneous equation in valence space. We use the Sternheimer~\cite{Sternheimer1950} or Dalgarno-Lewis \cite{Dalgarno1955} method implemented in the CI+all-order approach~\cite{Kozlov1999}. Given the $\Phi_0$ wave function and energy $E_0$ of the $|\Phi_0 \rangle$ state, we find intermediate-state wave functions $\delta \psi_{\pm}$ from an inhomogeneous equation,
\begin{eqnarray}
|\delta \psi_{\pm} \rangle & = & \frac{1}{H_{\rm eff} - E_0 \pm \omega}\,
 \sum_k | \Phi_k \rangle \langle \Phi_k | D_0 | \Phi_0 \rangle \nonumber \\
&=&  \frac{1}{H_{\rm eff}- E_0 \pm \omega} \, D_0 | \Phi_0 \rangle .
\label{delpsi}
\end{eqnarray}
Using Eq.~(\ref{Eqn_alpha}) and $\delta \psi_{\pm}$ introduced above, we obtain:
\begin{equation}
\alpha_v (\omega ) = \langle \Phi_0 |D_0| \delta \psi_+ \rangle
+ \langle \Phi_0 |D_0| \delta \psi_- \rangle  \, ,
\label{alpha2}
\end{equation}
where the subscript $v$ emphasizes that only excitations of the valence electrons are included in the intermediate-state wave functions $\delta \psi_{\pm}$ due to the presence of $H_{\rm eff}$.

The $\alpha_{c}$ and $\alpha_{vc}$ terms were evaluated in the RPA. The small $\alpha_{vc}$ term was calculated by adding $\alpha_{vc}$ contributions from the individual electrons. For example, for the $6s^2 6p^2 \,\, ^3\!P_1$ state, we find $\alpha_{vc} = 2 \alpha_{vc}(6s)+ \alpha_{vc}(6p_{1/2}) + \alpha_{vc}(6p_{3/2})$.

For the case of static polarizabilities, where $\omega=0$, Eq.~(\ref{Eqn_alpha}) is written as:
\begin{eqnarray}
\alpha_v (0) &=& 2\, \sum_k \frac {|\langle \Phi_0 |D_0| \Phi_k \rangle|^2} {E_k-E_0}.
\label{stat_alpha}
\end{eqnarray}

To establish the dominant contributions to the valence polarizabilities, we combine the electric-dipole matrix elements and energies according to the sum-over-states formula given by Eq.~(\ref{stat_alpha}). We have carried out two calculations of the dominant contributions of the intermediate states to the polarizabilities using our theoretical and experimental energies. In Table~\ref{tab_scal}, we present results obtained in the CI+all-order approximation. The absolute {\it ab initio} values of the corresponding reduced electric-dipole matrix elements are listed (in a.u.) in column labeled ``$D$.'' The theoretical and experimental \cite{NIST} transition energies are given in columns $\Delta E_{\rm th}$ and $\Delta E_{\rm expt}$. The remaining valence contributions are given in rows labeled ``Other.'' The contributions from the core and $\alpha_{vc}$ terms are listed together in the row labeled ``Core + Vc.'' The dominant contributions to $\alpha_0$, listed in columns $\alpha_0[\text{A}]$ and $\alpha_0[\text{B}]$, are calculated with CI+all-order+RPA matrix elements and theoretical [A] and experimental [B] energies \cite{NIST}, respectively. The results listed in the column $\alpha_0[\text{A}]$ are the recommended ones.

The results obtained in the CI+MBPT and CI+all-order approximations, their differences, and the recommended values are presented in Table~\ref{tab_polar}.

\section{Comparison of Experiment to Theory}
\label{discussion}

We turn now to the connection between our unitless $E2$/$M1$ amplitude ratio, $\chi$, and the theoretical expressions for the respective matrix elements. It is helpful to recall that both the $M1$ and $E2$ matrix element components emerge from the same term in the expansion of the interaction Hamiltonian. Following a standard textbook derivation of these higher-order terms \cite{Fitzpatrick}, we find that both the $M1$ and $E2$ transition amplitudes originate from a matrix element, $T_{fi}$, containing both the position and momentum operators,
\begin{equation}
\label{eqmultipole}
T _{fi \; (\rm{M1, E2})} \propto  \langle f | (\hat{k}\cdot\ {\bf r})\;(\hat{\epsilon}\cdot\ {\bf p} ) | i \rangle,
\end{equation}
where $\hat{k}~(\hat{z}$ in our case) is the laser propagation direction, and $\hat{\epsilon}~(\hat{x}$ in our case) is the laser polarization axis. We can ignore overall multiplicative factors since they will cancel in the eventual $E2$/$M1$ amplitude ratio.

After some vector algebra and use of a commutator to re-express the momentum operator in terms of position \cite{Fitzpatrick}, we can separate the $M1$ (vector) and $E2$ (second-rank tensor) components of the matrix element in Eq.~(\ref{eqmultipole}). We note that this process introduces a factor of $\omega_{\mathrm{E2}}/c$ into the $E2$ component. In our case, the $M1$ final state of interest is $|^3\!P_1, m=1 \rangle$, where as for the $E2$ component it will be $|^3\!P_2, m=1\rangle$.

According to the Wigner-Eckart theorem, for the case of the $|J=0\rangle \rightarrow |J_f, m=1\rangle$ transitions, the multiplicative factor connecting the $|\Delta m| = 1$ matrix elements that we measure with the associated {\it reduced} matrix element is $1/\sqrt{2J_f+1}$. Given our geometry, the operator for the $E2$ term is proportional to $\langle xz  \rangle$. This can then be rewritten in terms of the operator $\langle r^2 C_{21}\rangle$ as introduced in Sec.~\ref{background}.

Assembling a theoretical expression that is equivalent to the (unitless) experimental amplitude ratio $\chi$, given by Eq.~(\ref{Eqfinal}), we arrive at
\begin{equation}
\chi = \frac{1}{2\sqrt{5}} \frac{\omega_{E2}}{c} \, \frac{\langle ^3\!P_2 ||Q|| ^3\!P_0 \rangle}{\langle ^3\!P_1 ||\mu|| ^3\!P_0 \rangle}.
\label{chi_frac}
\end{equation}

Here the reduced ME $\langle ^3\!P_2 ||Q|| ^3\!P_0 \rangle$ is expressed in $|e| a_B^2$ (where $a_B$ is the Bohr radius; note that for this ME $1 \, {\rm a.u.} = 1\, |e| a_B^2$) and $\langle ^3\!P_1 ||\mu|| ^3\!P_0 \rangle$ is expressed in $\mu_B/c$.

Inserting our experimental value, $\chi = 0.1496(15)$, as well as the (highly accurate) theoretical value for the $M1$ reduced matrix element $\langle ^3\!P_1 ||\mu|| ^3\!P_0 \rangle = 1.293(1)\,\mu_B/c$~\cite{Porsev2016}, we can compute an experimentally-derived value for the reduced quadrupole matrix element: $\langle ^3\!P_2 ||Q|| ^3\!P_0\rangle_{\mathrm{exp}} = 8.91(9)$ a.u.. This is in excellent agreement with, and of comparable precision to, the recommended \emph{ab initio} theory value from Eq.~(\ref{eq:Qfinal}) in Sec.~\ref{theory}: $\langle ^3\!P_2 ||Q|| ^3\!P_0\rangle_{\mathrm{th}} = 8.88(5)$ a.u.. Together, we have demonstrated consistency between experiment and theory for this lead E2 transition amplitude at the 1.2\% level of accuracy.

\section{Concluding Remarks}
\label{conclusion}
We have completed a precise measurement of the electric quadrupole $^3\!P_0 \rightarrow \, ^3\!P_2$ transition amplitude within the $6s^2 6p^2$ configuration in atomic lead. This result is in excellent agreement with a precise \emph{ab initio} calculation of this amplitude, which has also been presented here. The calculation builds upon on recent theoretical work in the four-valence lead system aimed at improving PNC calculations in this element \cite{Porsev2016}. The experimental work relies critically on a high-precision polarimetry technique used previously to measure PNC optical rotation in Pb and Tl \cite{Meekhof1993, Vetter1994}, and has allowed direct measurement of this forbidden $E2$ transition for the first time.  We have also presented \emph{ab initio} calculations of the static polarizability of several low-lying states in lead.  This now provides additional opportunities to test the accuracy and further guide the refinement of theory through precise atomic-beam-based measurements of Stark shifts in this element, employing experimental techniques analogous to those used by our group in recent indium and thallium polarizability measurements \cite{Doret2002, Ranjit2013, Augenbraun2016, Vilas2018}.

\section*{Acknowledgments}

The authors thank Gabriel Patenotte and Sameer Khanbhai for their assistance in the construction and testing of the experimental apparatus. We are grateful for valuable conversations with David DeMille, and thank Charles Doret for helpful comments on the manuscript.  The experimental work described here was completed with the support of the National Science Foundation RUI program, through Grant No. PHY-1404206. The theoretical work was supported in part by NSF Grant No. PHY-1620687. S.~P.~acknowledges support by the Russian Science Foundation under Grant No. 19-12-00157.

\end{document}